\documentclass[prb,aps,epsf,twocolumn]{revtex4-2}
\usepackage{graphicx}
\usepackage{dcolumn}
\usepackage{bm}
\usepackage{color}
\usepackage{soul}
\usepackage{hyperref}
\usepackage{amssymb}
\usepackage{amsmath}
\usepackage{wasysym}
\usepackage{tabularx}

\usepackage{hyperref}

\begin{document}

\preprint{APS/123-QED}

\title{Phases and criticality of the triangular lattice $\text{SU}(N)$ Hofstadter-Hubbard model}

\author{Lu Zhang}
\affiliation{%
 Department of Physics, Hong Kong University of Science and Technology, Clear Water Bay, Hong Kong, China
}%

\author{Rongning Liu}
\affiliation{%
 Department of Physics, Hong Kong University of Science and Technology, Clear Water Bay, Hong Kong, China
}%
\author{Xue-Yang Song}
\affiliation{Department of Physics, Hong Kong University of Science and Technology, Clear Water Bay, Hong Kong, China} 
\date{\today}

\begin{abstract} 
    We report the study of phases and transitions of $\text{SU}(N)$ Hofstadter-Hubbard model subject to commensurate magnetic field on the triangular lattice. 
   At filling one fermion per site, for the number of fermion flavors $2\leq N\leq 8$, we identify three distinct phases and calculate critical interaction strength from parton large-$N$ mean-field approximation. Integer quantum Hall, chiral spin liquid, and valence bond solid states could be realized upon varying the Hubbard interaction $U$ and the number of flavor $N$.  We construct the critical theory for the putative continuous transition from quantum Hall states to chiral spin liquid and calculate the critical transport behavior using quantum Boltzmann equations for general $N$.
   These results could be validated in synthetic systems such as moir\'e superlattices and cold atom platforms. 
\end{abstract}

\maketitle

\section{Introduction}
Quantum spin liquid (QSL) has been proposed as the parent state of high-temperature superconductors~\cite{anderson1973resonating,savary2016quantum}.
It refers to highly entangled ground states for quantum magnets with emergent gauge structure and fractional excitations~\cite{savary2016quantum,zhou2017quantum,lee2008end,balents2010spin}. The search for quantum spin liquids and the transition into proximate phases has garnered significnat interests in 
condensed matter physics. 
Among various spin liquid states, chiral spin liquid(CSL) refers to a spin liquid state with non-vanishing spin chirality~\cite{wenchiral}. It is a gapped phase and has been searched for its topological nature and connection to anyon superconductivity~\cite{laughlin_anyon}. 
There have been various toy models proposed to host the CSL phases~\cite{he2014chiral,yao2007exact,schroeter2007spin,bauer2014chiral}. 
In particular, the $\text{SU}(N)$ Hubbard model is shown to host CSL from the parton mean-field approach~\cite{hermele2009mott,chen2016synthetic,yao2021topological}. 
Recently, numerical studies~\cite{divic2024chiral,ding2024particle,kuhlenkamp2024chiral,divic2024chiral,kuhlenkamp2024aspects} on the Hofstadter-Hubbard model on the triangular lattice proposed various magnetic and topological phases. 
It is also demonstrated that the anyon superconductivity can be obtained by doping the model near the critical point~\cite{divic2024anyon}.

\begin{figure*}
\includegraphics[width=.8\linewidth]{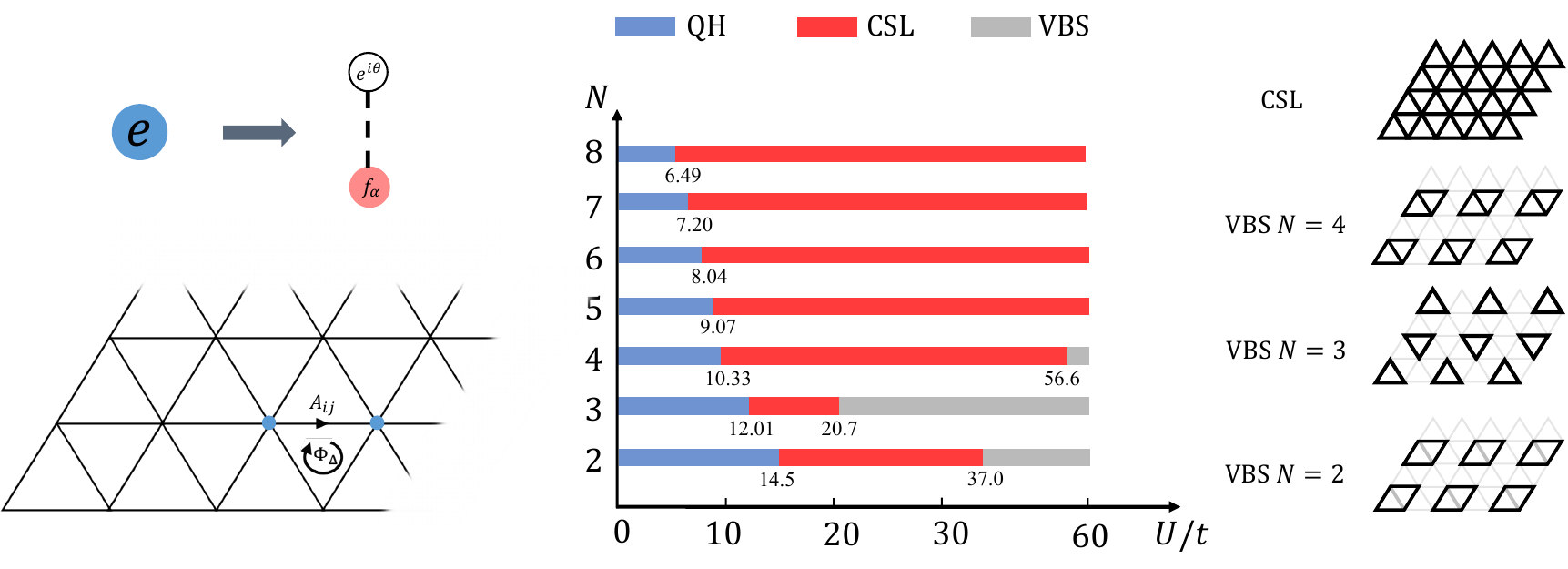}
    \caption{Left: the triangular lattice with gauge field $A_{ij}$ defined on the bond. Flux $\Phi_\Delta$ piercing each triangle is $\pi/N$. The electrons feel an on-site Hubbard interaction $U$, which is fractionalized into the rotor and spinon degrees of freedom. In the large $U$ limit, the rotor is gapped. Middle: phase diagram of the Hofstadter-Hubbard model of coupling strength $U/t$ and the $N$. It is calculated using the slave rotor mean-field in the weak coupling limit and spinon parton in the strong coupling limit. Starting from weak interaction limit, the system hosts IQH states(blue) and transitions into CSL (red) as $U/t$ increases. The grey region indicates VBS phases. The critical points $U_{c_1}$ and $U_{c_2}$ are the critical interaction strength for IQH-CSL and CSL-VBS transition, respectively. Right: the schematic representation of the ansatz $\chi_{ij} = \sum_{\alpha}\langle f_{i\alpha}^\dagger f_{j\alpha} \rangle/N$ of chiral spin liquid(CSL) states, valence bond solid(VBS) as a function of $N$. The thickness of the link schematically represents the magnitude of $\chi_{ij}$.}
    \label{fig:main}
\end{figure*}

In this paper, we systematically investigate the $\text{SU}(N)$ Hofstadter-Hubbard model on the triangular lattice(See Fig.\ \ref{fig:main}) with one electron at each site.
This model is the $\text{SU}(N)$ generalization of  the $\text{SU}(2)$ Hofstadter-Hubbard model. In contrast to the $\text{SU}(N)$ Hubbard model, there exists a finite magnetic flux threading each plaquette. The magnetic field couples to the orbital degrees of freedom through minimal coupling. The Zeeman coupling is omitted as the spin flavors in $SU(N)$ originate from enhanced pseudo-spin symmetry, e.g. layer degrees of freedom, in moire or synthetic platforms. From perturbation theory, this flux leads to a spin chirality term $e^{i\Phi_{\triangle}}\mathbf{S}_i \cdot\left(\mathbf{S}_j \times \mathbf{S}_k\right)$, which penalizes the ordered phase and favors the CSL energetically~\cite{hu2016variational,wietek2017chiral}. 
Recently, it is numerically demonstrated that the ground state of $\text{SU}(2)$ Hofstadter-Hubbard model is the $\text{U}(1)$ CSL in a wide range of interaction parameters and there exists a continuous phase transition from CSL to integer quantum Hall phase(IQH)~\cite{divic2024chiral}.  While it is  computationally costly to study the model for $N\geq 2$, we adopt the controllable large-$N$ parton mean-field approach to study the ground state of $\text{SU}(N)$ Hofstadter Hubbard model,which becomes reliable in the large-$N$ limit.
As $U$ increases, the electrons tend to be localized by the strong on-site repulsion.
Our calculation reports the competition between the valence bond solid(VBS) and the CSL, where we use VBS to refer to all the translation symmetry breaking phase including the plaquette state and dimerized state in this paper. 
It is found in the mean-field calculation that the CSL can be stabilized in a wide parameter region while the CSL is absent for the pristine $\text{SU}(N)$ Hubbard model, i.e. without external magnetic field, when $N\leq 5$.  In the weak coupling limit, the system enters an integer quantum Hall(IQH) phase. The competition between the IQH and the CSL is captured by the rotor formalism.

In addition to the topological phases in different coupling regimes, this model realizes the quantum critical point of the IQH to CSL transition. The putative continuous phase transition can be described by the scalar $\text{QED}_3$ coupled with Chern-Simons term~\cite{divic2024anyon} with the level determined by $N$. As an interacting theory proposed to be conformally invariant~\cite{chester2018monopole}, it is of theoretical significance to study the universal transport properties, as can be obtained by quantum Boltzmann equation.  It is observed that the longitudinal conductivity jumps to finite value at the critical point while stays zero deep in the topological phases on both sides and the transverse conductivity jumps from finite to zero as the system transits from IQH to CSL phase, a feature shared in the proposed continuous transition out of fractional quantum anomalous Hall phase in moire systems\cite{song_2024_phase}.

This model is not only of conceptual importance but also experimentally inspired. When $N=2$ and $N=4$, the model can be potentially realized in moir\'e bilayers and double moir\'e bilayers under a magnetic field, respectively, by treating the layer degrees of freedom as the pseudo-spin~\cite{kuhlenkamp2024chiral,zhang2021su4}. The more general $\text{SU}(N)$ Hofstadter-Hubbard model can also be modeled in fermionic alkaline-earth atomic systems with synthetic gauge field~\cite{yang2024chiral,heinz2020state,aidelsburger2013realization,cooper2019topological,gorshkov2010two,honerkamp2004ultracold,rapp2008trionic,wu2003exact}. There are two essential features that enhance the symmetry of the system to $\text{SU}(N)$:~(i) the large nuclear spin $I$ and (ii) the decoupling of the nuclear spin degrees of freedom from the electronic state in ground state (${ }^1 S_0$) and first excited states(${ }^3 P_0$). Combining these features, the $\text{SU}(2I+1)$-symmetric interaction can be effectively simulated in a controllable environment. Hence the model bears important relevance to quantum simulator platforms.

\section{Main results}
\label{sec:Hofstadter Hubbard model}
The $\text{SU}(N)$ Hofstadter-Hubbard model on the triangular lattice is defined as
\begin{equation}
H = -t\sum_{\langle  i 
 j \rangle \alpha} \left( e^{-i\bar A_{ij}} c^\dagger_{i\alpha}c_{j\alpha} +\text{H.c.}\right)
 +\frac{U}{2}\sum_{i,\alpha\neq \beta}n_{i\alpha}n_{i\beta},
 \label{eq:Hubbard}
\end{equation}
where $\alpha = 1,\cdots, N$ is the flavor indices and $\bar A_{ij}$ is the external electromagnetic gauge field with flux $\Phi_\triangle = \pi/N$ threading each triangular plaquette. In the presence of the magnetic flux, the unit cell must be enlarged $N$ times and the Brillouin zone is reduced to the magnetic Brillouin zone correspondingly, which leads to the band folding. We focus on the filling of one electron per site. The degrees of freedom at each site form a fundamental representation of the $\text{SU}(N)$ flavor symmetry. 

To describe the low energy properties of the model,
we adopt the parton formalism by decomposing the electron as $c^\dagger_{i\alpha} = e^{i\theta_i}f_{i\alpha}^\dagger$
with $f^\dagger_{i\alpha}$ creating a spinon and $e^{i\theta_i}$ annihilating a rotor excitation at each site.
We use the following convention: when each site is singly occupied by an electron, rotor occupation, i.e. the angular momentum, is zero and the spinon occupation is one. This parton decomposition introduces gauge redundancy as the physical electron operator is invariant under a phase rotation $e^{i\theta_i}\rightarrow e^{i\theta_i+i\phi},f_{i\alpha}^\dagger\rightarrow f_{i\alpha}^\dagger e^{-i\phi}$. Hence one introduces a $U(1)$ gauge field $a$ to account for the gauge fluctuations at low energy. The rotor and spinon both carry $+1$ charge under $a$.
It is known that the strong Hubbard repulsion suppresses the double occupancy. Therefore, the charge dynamics are quenched, and  rotor field is gapped out in the large-$U$ limit.
The spinons hence capture
the low energy dynamics. 
Specifically, $\text{U}(1)_N$ CSL corresponds to the $C=N$ Chern insulator of spinons. The translation breaking VBS state corresponds to a spinon state with modulated bond strength that represents formation of valence bonds or plaquettes.

The rotor $e^{i\theta_i}$ starts to play a role when $U$ decreases. The formation of IQH corresponds to the condensation of rotor fields i.e. $\langle e^{-i\theta_i}\rangle\neq 0$, which relates  the electrons to spinons from $c_{i\alpha}=\langle e^{-i\theta_i}\rangle f_{i\alpha}$. We will discuss the transition given by the rotor condensation process.
In the limit $U=0$, the Hamiltonian can be exactly solved with the $C=N$ IQH phase as the ground state. 
The main result of our study is shown in the Fig.~\ref{fig:main}. The above results are supported by a self-consistent parton mean-field calculation covering a wide range of coupling strength $U/t$. 

Next we elaborate on the methodology and phase diagrams. In section \ref{sec:strong_coupling} we elaborate on the strong coupling (large $U/t$) regime where competition between CSL and translation breaking phase is resolved by spinon mean-field theory. In section \ref{sec:QH to CSL phase transition} we discuss the weak interaction regime with the IQH phase in the free electron limit, and compute the critical interaction strength for transition out of the IQH state, based on the rotor mean-field theory. In both sections we will supplement the discussion with physical arguments and comparison with well-studied models such as Hubbard and Heisenberg models, which puts our mean-field results on a firmer footing. In section \ref{sec:The critical theory between IQH and CSL} we discuss the transition from IQH to CSL and the critical transport properties of the putative continuous transition from IQH to CSL.


\section{CSL and VBS phases in Strong Coupling Regime\label{sec:CSL in Hofstadter}}
\label{sec:strong_coupling}
In this section, we describe the CSL and VBS phase in the strong coupling regime, in which the rotor field is gapped out.
We apply the perturbation theory to reduce the Hubbard model to the $J-K$ model:
\begin{equation}
    H = J\sum_{\langle ij \rangle} S^{\alpha \beta}_i S^{\beta \alpha}_j + K \sum_{\langle ijk\rangle} e^{-i\Phi_{ijk}}S_i^{\alpha\beta} S_j^{\beta \gamma} S_k^{\gamma \alpha},
    \label{eq:JK}
\end{equation}
where $S^{\alpha \beta}$ is the $\text{SU}(N)$ generators and the Einstein summation is adopted for repeated Greek symbols. The parameters $J = \frac{2t^2}{U}$ and $K = \frac{6t^3}{U^2}$ are determined from the perturbation theory in Appendix~\ref{appendix:tJ model}.  The notations '$\langle ij \rangle$' and '$\langle ijk \rangle$' represent that the two sites $i,j$ are nearest connected and the three sites $i,j,k$ belong to the same triangle(including the upward-pointing and downward-pointing triangles) in the lattice. 
Written in terms of the spinon creation and annihilation operators $S_i^{\alpha \beta} = f_{i \alpha}^{\dagger} f_{i \beta}$, the $J-K$ model can be expressed as
\begin{equation}
\begin{aligned}
    H &= -J\sum_{\langle ij\rangle} f^\dagger_{i\alpha}f_{j\alpha}f^\dagger_{j\beta}f_{i\beta} \\&~~~~~~~+K\sum_{\langle ijk\rangle}e^{-i\Phi_{ijk}}f^\dagger_{i\alpha}f_{j\alpha}f^\dagger_{j\beta}f_{k\beta}f^\dagger_{k\gamma}f_{i\gamma}.
\end{aligned}
\label{eq:parton_decomposition}
\end{equation}
We impose that each site contains only one spinon
$\sum_\alpha f^\dagger_{i\alpha} f_{i\alpha} = 1$
to project to the physical Hilbert space.  This can be accomplished by introducing the Lagrangian multiplier $\mu_i$.  A more detailed discussion can be found in section II of the Ref.~\cite{wen2002quantum}. 
In the following we adopt the standard large-$N$ scheme by keeping $\mathcal J = NJ$ and $\mathcal K = N^2 K$ constant. The following mean-field Hamiltonian can be obtained by variational principle~\cite{brinckmann2001renormalized}:
\begin{equation}
\begin{aligned}
H &= -\mathcal J\sum_{\langle ji \rangle} \chi_{ji} f^\dagger_{i\alpha}f_{j\alpha}\\
&~~~~~~+  \mathcal K\sum_{\langle ijk\rangle}e^{-i\Phi_{ijk}}\chi_{ij}\chi_{jk}f^\dagger_{k\alpha}f_{i\alpha}-\sum_i \mu_i f^\dagger_{i\alpha} f_{i\alpha}
\end{aligned}
\label{eq:SCequation}
\end{equation}
where $\chi_{ij} \equiv \sum_{\alpha}\langle f^\dagger_{i\alpha} f_{j\alpha} \rangle/N$ is the mean-field parameter, which can be identified as the order parameter of the VBS and also diagnose chiral symmetry breaking in the CSL.
The gauge constraint can be satisfied on average by adjusting the chemical potential, such that
$\sum_\alpha\langle f^\dagger_{i\alpha}f_{i\alpha}\rangle = 1$.
We begin with a random initial ansatz $\chi_{ij}$ and $\mu_i$ and solve the Hamiltonian~\eqref{eq:SCequation} at the mean-field level self-consistently (details in Appendix~\ref{appendix:mean-field}).

The properties of ground states can be identified from the configuration of mean-field parameters, i.e. the bond strength $\chi_{ij}$. CSL corresponds to the uniform bond strength with flux piercing each plaquette, whereas the VBS corresponds to the translation symmetry breaking pattern of bond strength. In particular, the bond strength $\chi_{ij}$ is uniform in the chiral spin liquid phase with the $\arg(\chi_{ij}\chi_{jk}\chi_{ki})=\pi/N$ flux experienced by the spinons. This indicates the non-trivial topology in the system, as the mean-field band structure for the spinons furnishes a Chern band. The phase diagram is shown in the middle panel of Fig.~\ref{fig:main}. The intermediate interaction regime admits CSL as the ground state, while the system enters translation breaking phase, i.e. VBS, as $U$ further increases for $2\leq N\leq 4$. As $N$ increases, CSL phase becomes more stable (except for the special case $N=3$ which will be elaborated below) and for $N\geq 5$, the CSL becomes robust and is favored over translation breaking state even at strong interaction. The details of parton calculation and resulting mean-field states are explained in Appendix~\ref{appendix:mean-field}.

\begin{figure}
    \centering
    \includegraphics[width=1.0\linewidth]{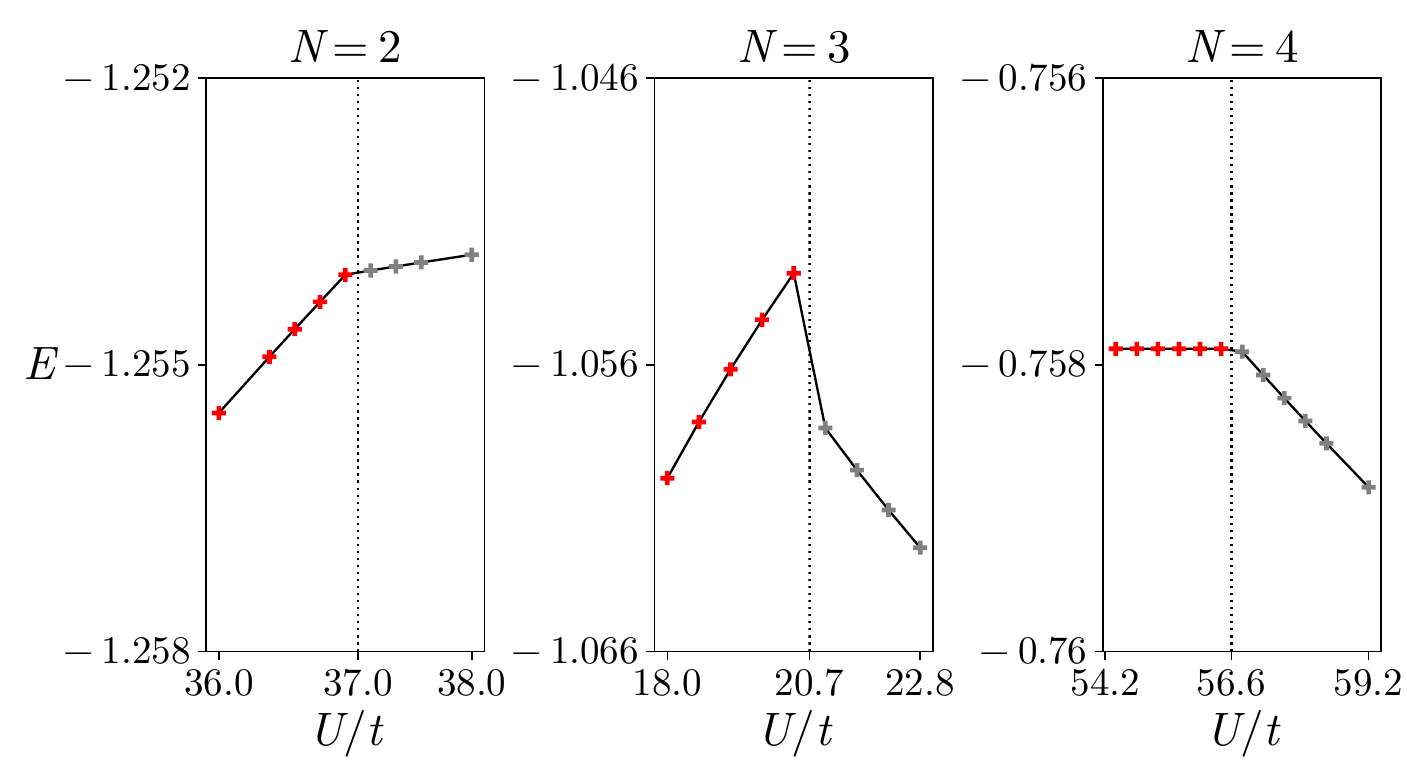}
    \caption{The ground state energy per site as a function of interaction strength $\frac{U}{t}$. The critical interaction strength is marked by the dotted vertical line. The red marker represents the phase of CSL and gray markers represent the VBS phase.  The energy is defined in the unit of $10N^2N_sJ$.}
    \label{fig:energy_CSLVBS}
\end{figure}


To understand the result, we start from $\text{SU}(N)$ Heisenberg model $J S_i^{\alpha \beta} S_j^{\beta \alpha}$, which is the first term in our $J-K$ model with coupling strength $J \sim 1/U$. When $N<5$, the ground state for the Heisenberg model breaks translation and forms the VBS~\cite{yao2021topological}, with CSL the closest competing state. To the next order perturbation, the external magnetic field introduces the chiral interaction $e^{-i \Phi_{i j k}} S_i^{\alpha \beta} S_j^{\beta \gamma} S_k^{\gamma \alpha}$, which energetically favors the CSL phase. The coupling strength of this chiral term is $K \sim \frac{1}{U^2}$.
In the limit of infinite $U$, $J \gg K $, the system is dominated by the Heisenberg term, which favors VBS over CSL~\footnote{Heisenberg term also favors magnetic order, and the magnetic order escapes the spinon decomposition adopted here, which will be discussed toward the end of the main text.}.
As Hubbard $U$ decreases, the chiral term becomes increasingly important according to the scaling of $J$ and $K$ over $U$. The system remains in the VBS phase until it reaches the critical point, $U_{c_2}$, below which the ground state transitions to the CSL. 

The ground state energy per site near the critical point is shown in the Fig.~\ref{fig:energy_CSLVBS}, which is used to determine the critical $U_{c2}$.  For $\text{SU}(2)$ Heisenberg model any configuration of dimers that fully covers the lattice has exactly the same energy on the mean-field level. The superposition of any dimer states is one of the mean-field solution. It is believed that one of the dimerized states will be selected upon considering $1/N$ corrections. With the chiral term, the plaquette order should be more energetically favored than the fully dimerized state. 
When $U \rightarrow  \infty$, the result is consistent with the previous study~\cite{yao2021topological} in the unit $N^2N_sJ$ ($N_s$ is the number of total lattice sites). 

When $N\ge5$, there is no translation symmetry breaking phase and ground state stays as CSL throughout large $U$ regime. 
This finding is also natural given that the VBS phase already ceases to be the ground state of the Heisenberg model at $N>5$~\cite{hermele2009mott,yao2021topological}. 
When $N=5$ the energy of the VBS(stripe order) and the energy of CSL are almost identical such that the chiral term always leads to the CSL phase in the coupling regime shown in our phase diagram~\ref{fig:main}.

An anomaly is that the $U_{c_2}$ decreases for $N=3$, compared with $N=2$. 
This is due to the energy cost of triangle plaquette in the $N=3$ VBS is smaller than in $N=2,4$ cases.  Hence VBS is favored in $\text{SU}(3)$ case. It is also observed that there is an energy jump at the phase transition from CSL to VBS when $N=3$, shown in Fig.~\ref{fig:energy_CSLVBS}. 
It is worth noting that mean-field result ignores the quantum fluctuations and hence the nature of the phase transition is left for the future study.

\section{IQH and CSL at weak interactions}
\label{sec:QH to CSL phase transition}
To study the weak interaction regime, which covers the IQH and the phase transition IQH-CSL, we resort to the original complete parton construction.
The $J-K$ model used in strong coupling regime, derived from the perturbation theory, is based on the fact that there is no double occupancy when onsite interaction $U$ is strong.
However, when the interaction strength is not strong enough to quench the charge dynamics, the perturbation theory breaks down.  

To describe the low-energy physics in the weak-coupling regime, we have to include the rotor variable by expressing electron creation operator as the combination of spinon and rotor on each site, i.e., $c^\dagger_{i\alpha} = e^{i\theta_i}f_{i\alpha}^\dagger$(See Sec.~\ref{sec:Hofstadter Hubbard model}). 
The gauge constraint
$
L_i + \sum_{\alpha}f_{i\alpha}^\dagger f_{i\alpha}= 1
$
has to be imposed, where $L_i = i\partial_{\theta_i}$ is the conjugate variable to $\theta_i$. The eigenvalue of $L_i$ corresponds to the occupation of rotor. According to our definition, the rotor number $L_i = 0$ corresponds to filling one electron per site. When the filling deviates from one electron per site, the occupation number of rotors measured by $L_i$ in non-zero. This should lead to the energy penalty due to the Hubbard interaction. Formally, the particle number operator can be written as $n_i=\sum_\alpha f_{i,\alpha}^\dagger f_{i,\alpha}=1-L_i$. Hubbard term can be written as $\sum_{\alpha\neq \beta}f^\dagger_{i\alpha}f_{i\alpha}f^\dagger_{i\beta}f_{i\beta}$.  Replacing density operator $n_i$ in the Hubbard term with $L_i$ leads to
\begin{equation}
\begin{aligned}
    H = &-\sum_{\langle ij \rangle\alpha}te^{-i\bar A_{ij}}e^{-i(\theta_i - \theta_j)}f_{i\alpha}^\dagger f_{j\alpha}
    \\& ~~~~~ +\frac{U}{2}\sum_i L_i^2 +
    \sum_i\mu_i\left(L_i+\sum_\alpha f^\dagger_{i\alpha}f_{i\alpha}\right) ,
    \label{eq:H_slave rotor}
\end{aligned}
\end{equation}
where $\mu_i$ is introduced to force the gauge constraint $L_i+\sum_\alpha f^\dagger_{i,\alpha}f_{i,\alpha} = 1$. In this formulation, the Hubbard interaction only depends on the rotor degrees of freedom.
 The problem can be formulated in the Euclidean path integral 
 $$\mathcal Z = \int \mathcal D[\psi, L, f,\lambda,\mu] \delta\left(|\psi|^2-1\right)\exp\left[H(\psi,f,L,\mu)+\sum_iL_i\partial_\tau\psi_i \right]$$
Integrating out the conjugate variable $L$ leads to the action:
\begin{equation}
\begin{aligned}
    &\mathcal S = \int_\tau  \frac{1}{2U}\sum_{i}|(\partial_\tau -\mu_i)\psi_i|^2-\sum_i\lambda_i(|\psi_i|^2-1)  \\
    &~~~~~+\sum_{i,\alpha}f^\dagger_{i\alpha} (\partial_\tau-\mu_i) f_{i\alpha}-  t \sum_{\langle ij \rangle,\alpha}e^{-i\bar A_{ij}}\psi_i^\dagger \psi_j f_{i\alpha}^\dagger f_{j\alpha},
\end{aligned}
\label{eq:partition_function}
\end{equation}
where $\psi$ is the rotor field $\psi_i \equiv e^{i\theta_i}$ and $\lambda_i$ is the Lagrangian multiplier to enforce the constraint of field $|\psi_i|^2=1$.  

The saddle point of the above action leads to a set of equations which is solved self-consistently. We can hence determine the dispersion of the spinon and rotor fields (see Supplemental Material). The spinons fill Chern bands as in the previous section. The critical interaction $U_{c_1}$(shown in Fig.\ \ref{fig:main}) is fixed by the condition that the minimal energy of the rotor hits zero, leading to rotor condensation. This condensation restores phase coherence among
the rotors, which formally identifies the dynamics of electrons $c_{i,\alpha}$ with those of the spinons $f_{i,\alpha}$. This leads to the
emergence of an IQH state as the electrons will fill Chern bands with $C = N$ which descends from the spinon Chern
bands.  
 When $N=2$, the critical point $U_{c_1}/t \approx 14.5$, close to the numerical result found by DMRG $U_{c_1}/t \approx 12$~\cite{divic2024chiral}. As $N$ increases, the CSL becomes more robust, while the IQH phase shrinks with respect to $U/t$, in line with the observation that flatter Hofstadter band enhances correlation effects that favors CSL.

We note that this approach naturally implies that the IQH-CSL transition is second order at mean-field level and controlled by the condensation of the rotor fields\cite{divic2024chiral,divic2024anyon}. Next we study closely this putative continuous transition beyond mean field.

\section{The criticality between IQH and CSL}
\label{sec:The critical theory between IQH and CSL}
In this section, we go beyond the mean-field description of the phase transition and establish the critical theory near the critical point between IQH phase and CSL phase. The gauge field $\alpha_{ij}$ emerges when we consider the fluctuation of the bond strength, i.e. $\chi_{ij}\equiv t e^{-i\bar A_{ij}} \langle f_{i\alpha}^\dagger f_{j\alpha}\rangle$ and in general with fluctuations $\chi_{ij} = |\chi_{ij}|e^{i\alpha_{ij}}$. We start by coarse-graining the action~\eqref{eq:partition_function}.  In real space the hopping of the rotor is identified with the $\partial_x^2+\partial_y^2$ in continuum limit. The condensation of the rotor is described by the Lagrangian,
\begin{equation}
\begin{aligned}
        \mathcal{L}^r[\phi,a,A,\lambda] =& \frac{1}{g}|(\partial - ia+iA)\phi|^2 + \frac{i\lambda}{g}(|\phi|^2-1)+\cdots,      
\end{aligned}
\label{eq:rotor_condense}
\end{equation}
where $\phi$ represents the rotor field in the long wave length and satisfied the constraint $|\phi(x)|^2=1$ by integrating out the $\lambda$. The higher order irrelevant terms are neglected in the Lagrangian~\eqref{eq:rotor_condense}.  

By choosing the proper unit of space and time, the coefficients for $\partial^2$ and $\partial_\tau^2$ terms can be set equal and the coupling $g \propto \frac{U}{t}$. The coupling of the Lagrange multiplier term $\lambda (|\phi|^2-1)$ is chosen to be $1/g$ for convenience of the future calculation of conductivities, by scaling the field $\lambda$~\footnote{This rescaling does not change the result since this term only enforces the unit constraint.}.

Since the spinon is in the integer quantum Hall phase, integrating out the spinon field gives rise to the Chern-Simons term. We have the following effective theory for the spinons:
\begin{equation}
    \begin{aligned}
        \mathcal{L}^s[\alpha_i,a] =& -\frac{1}{4\pi}\sum_i^N\left(\alpha_id\alpha_i+ad\alpha_i\right) ,\\
\end{aligned}
\end{equation}
where $\alpha_i$ is the dual gauge field of the spinon $3$-currents, $a$ the internal gauge field from the parton decomposition and $A$ is the electromagnetic field coupling to the  electron~\footnote{In the following we neglect the response of the flavor degrees of freedom thus we do not include the spin gauge field $A_s$.}.
The complete critical theory consists of the above two parts, 
\begin{equation}
\begin{aligned}
        \mathcal{L} =& \mathcal{L}^r[\phi,a,A,\lambda] + \mathcal{L}^s[\alpha_i,a].
\end{aligned}
\end{equation}
Further integrating out $\alpha_i$'s, we are left with the rotor degrees of freedom.
Then we arrive at the critical theory for  the phase transition between the CSL and IQH:
\begin{equation}
    \mathcal L= \frac{1}{g}|(\partial -ia+iA)\phi|^2+ \frac{i\lambda}{g}(|\phi|^2-1)+\frac{N}{4\pi}ada+\cdots,
    \label{eq:critical boson}
\end{equation}
where the $\cdots$ represent the Maxwell term in our effective action that is irrelevant in the presence of the Chern-Simons term. 

IQH and CSL phases are separated by the critical point. In the IQH($\delta \equiv g_c^{-1}-g^{-1}<0$) phase the rotors spontaneously condense, which leads to the Higgs phase. The emergent gauge $a$ and the external gauge field $A$ are locked. The low energy physics is described by the following response action:
\begin{equation}
    \mathcal{L} = \frac{N}{4\pi} AdA.
\end{equation}
While in the CSL($\delta>0$) phase there is a charge gap. Then the rotor fields can be integrated out, which leads to the effective topological field theory of emergent gauge field $a$: 
\begin{equation}
\mathcal{L} = \frac{N}{4\pi} ada .
\end{equation}
The spinon gap remains open through the transition~\footnote{It should be noted that the gravitational Chern-Simons term should be added to reproduce the correct thermal Hall effect. But it is not relavent to this study.}. Thus the low energy physics is determined by the critical boson described by the Lagrangian~\eqref{eq:critical boson} near the phase transition point. 

In the following, we discuss the physical consequence of the critical theory which can be tested experimentally.
\begin{figure}
    \centering
    \includegraphics[width=.8\linewidth]{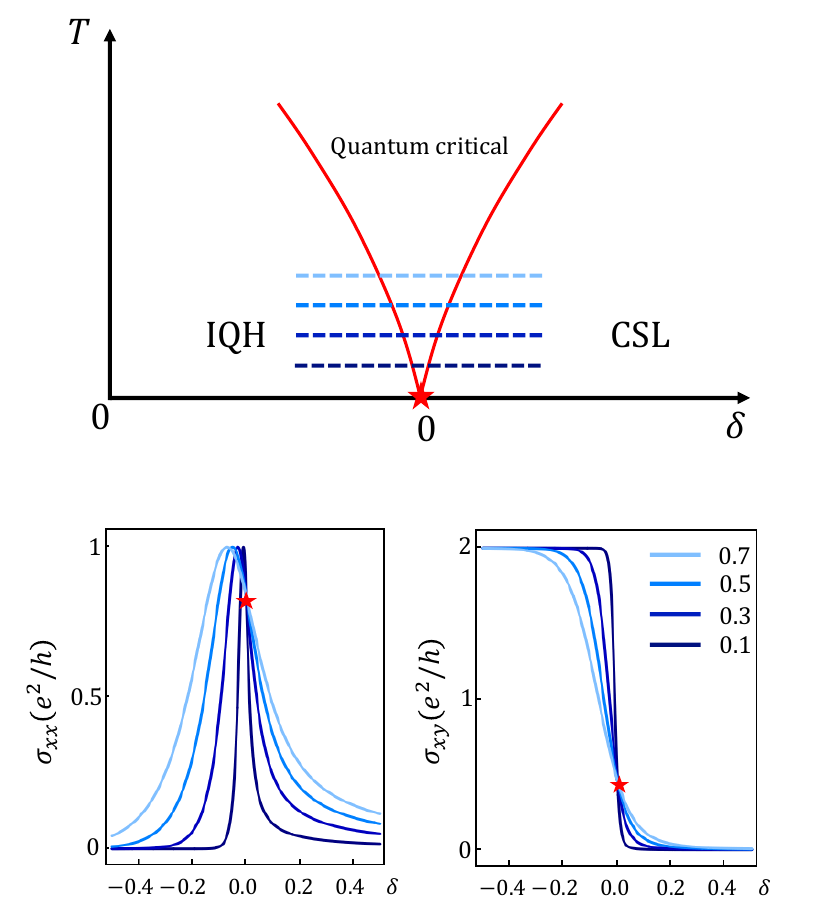}
    \caption{Top: schematic plot of the finite temperature phase diagram from the scaling theory ($\delta$ is the tuning parameter). The phase boundary(red line) between the quantum critical region and the topological phases is determined by $T = \delta^{\nu z}$. Bottom: the numerical result of the longitudinal(left) and transverse(right) conductivity when $N=2$ by tuning the parameter $\delta$ at various temperatures $T = 0.1,0.3,0.5,0.7$. The temperature $T$ and $\delta$ are both given in a common and arbitrary unit of energy.  In the quantum critical fan, the conductivity curve is universal and can be obtained from quantum Boltzmann equation.  }
    \label{fig:universal jump}
\end{figure}
The resistivity is given by the Ioffe-Larkin rule
\begin{equation}
\rho_e = \rho_\phi + \rho_{\text{CS}}.
\label{eq:CSRPA}
\end{equation}
In our case, the $\rho_\phi$ is the resistivity tensors of the critical boson and $\rho_{\text{CS}}$ can be expressed as
\begin{equation}
    \rho_{\text{CS}} = \left(\begin{array}{ll}
    0 & \rho_H \\
    -\rho_H & 0
\end{array}\right),
\end{equation}
where $\rho_H = \frac{1}{N}\frac{h}{e^2}$. This could be more rigorously derived by the Chern-Simons RPA approach, which can be understood in a pictorial way. In the presence of the Chern-Simons term, the flux is attached to the charge by integrating out the time component of the gauge field $a$. When there exists charge current moving along direction $x$, the flux is also flowing along with the charge current. Consequently there is a transversal electric field generated by moving flux according to the Faraday's law.
Given the electric current $\boldsymbol{j}$, the total electric field experienced by the charge consists of the external electric field and the electric field $\boldsymbol{E}$ generated by the Chern-Simons term $\rho_{\text{CS}}\boldsymbol{j}$ such that $\rho_\phi \boldsymbol{j} = \boldsymbol{E}-\rho_{\text{CS}}\boldsymbol{j}$. The more detailed illustration can be found in Ref.~\cite{simon1998chern}.

In the IQH phase, the holon condenses thus resistivity of holon vanishes $\rho_\phi=0$, while in the CSL phase, $\rho_\phi = +\infty$. At the critical point the resistivity of the slave rotor is $\rho_\phi^c = R\frac{h}{e^2}$ and $R$ can be evaluated numerically~\cite{witczak2012universal}. The resistivity for the critical rotors in Eq.~\eqref{eq:critical boson}, is a universal number intrinsic to the critical theory. Therefore, the total resistivity at the critical point should be
\begin{equation}
    \rho^c_e = \left(\begin{array}{ll}
    R\frac{h}{e^2} & \rho_H \\
    -\rho_H & R\frac{h}{e^2}
\end{array}\right).
\end{equation}
The quantities $R$ and $\rho_H$ appearing in the above formula are the universal numbers. Combining the above arguments we can obtain the conductivity behaviors shown in Fig.\ \ref{fig:universal jump}(see Supplemental materials) and predict there is a universal resistivity jump once the critical point is reached, which can serve as a hallmark signature for the topological criticality in transport measurements. The conductivity jump will be rounded at non-zero but low temperature in a universal manner, and follows a scaling form in the quantum critical fan
\begin{align}
 \sigma_{ij}(\delta, T) = \frac{e^2}{h} S_{ij}\left( \frac{\delta}{T^{\frac{1}{\nu z}}} \right)  
 \end{align}
 where $S_{ij}(x)$ is a universal $2 \times 2$ matrix, and $\nu,z$ are the correlation length critical exponent and dynamical exponent, respectively. The structure of the critical theory implies $z=1$.

The universal value $R$ and dependence of the resistivity on the temperature and tuning parameter can be calculated using the quantum Boltzmann equation~\cite{mahan2013many,damle1997nonzero,sachdev1998nonzero,witczak2012universal,xu2022interaction,lee1995quantum}:
\begin{equation}
\begin{aligned}
    \left(\frac{\partial}{\partial t}+s\boldsymbol{E} \frac{\partial}{\partial \boldsymbol{k}}\right)f_s(t,\boldsymbol{k}) = \frac{1}{2}I_\lambda(f_{s}(t,\boldsymbol{k})),
\end{aligned}
\end{equation}
where $f_s(t,\boldsymbol{k})$ is the distribution function of the holon, $s=\pm 1$ and the right-hand side $I_\lambda$ is the collision term (details in Supplemental materials). The quantum Boltzmann equation  is applicable to the hydrodynamic regime such that $\hbar\omega\ll k_BT$, where the dissipation is dominated by the incoherent scattering of the critical model. 
The solution of the quantum Boltzmann equation at different $N$ is shown in the table~\ref{tab:my_label}. 

\begin{table}
    \begin{center}
    \renewcommand\arraystretch{1.2}
    \begin{tabular}{|m{2em}|m{3em} m{3em}m{3em}m{3em}m{3em}m{3em}m{3em}|}
         \hline
        ~$N$ & ~2  & ~3 & ~4 & ~5 & ~6  & ~7 & ~8  \\ [0.2ex]
        \hline 
        $\sigma_{xx}$ & $0.840$ & $0.961$ & $1.013$ & $1.039$ & $1.053$ & $1.062$  & $1.068$   \\ [0.2ex] 
        \hline
        $\sigma_{xy}$ & $0.457$ & $0.348$ & $0.276$ & $0.226$ & $0.191$ & $0.165$ & $0.145$  \\ [0.2ex] 
        \hline 
    \end{tabular}
    \end{center}
    \caption{The longitudinal and transverse conductivities in the unit of $\frac{e^2}{h}$ at the critical point for different $N$.}
    \label{tab:my_label}
\end{table}


%
%
%

\section{Conclusion and Discussion}
To summarize, we have thoroughly studied the $\text{SU}(N)$ Hofstadter Hubbard model on the triangular lattice.
This paper demonstrates that in the strong coupling limit the ground state can host the chiral spin liquid, or valence bond solid by employing the spinon mean-field method. 
In the weak coupling limit, the ground state is the integer quantum Hall since the interaction strength is small compared with the band gap.
Additionally, we have determined the corresponding phase boundary between IQH and CSL via rotor condensation. 
By examining the critical theory proposed, we obtain a universal jump of conductivity from the IQH to CSL phase transition.
The universal conductivity intrinsic to the putative critical theory is obtained through the quantum Boltzmann equation. The phase diagrams and critical behaviors can be further checked by numerical calculations in the future. Our work established the $\text{SU}(N)$ Hofstadter-Hubbard model to host various interesting topological phases and revealed the properties of the topological criticality between IQH and CSL. This calls for the search of model realization in cold atom or solid state systems. In particular, the effects of psedo-spin symmetry breaking in microscopic systems should be carefully investigated. This model not only hosts exotic topological phases, but also potentially connects to anyon superconductivity with charge-$N$e (for even $N$) pairing by doping the critical point\cite{divic2024anyon}. Hence it is both of conceptual and practical importance to further study the model at small doping and its realization in solids or quantum simulators.

We comment on the limitation and justification of our methodology. 
Admittedly,the large-$N$ mean-field approach captures only partially the quantum fluctuations when $N$ is small. 
The previous studies have shown that the magnetic order is stabilized when $N=2,3$ in the large-$U$ limit of Hofstadter model~\cite{white2007neel,lauchli2006quadrupolar,bauer2012three}, which differs from the mean-field results of the fermionic parton theory. 
We employ fermionic spinon decomposition since it most naturally captures CSL, which is shown to be stabilized in the $\text{SU}(2)$ Hofstadter-Hubbard model in the intermediate coupling regime~\cite{divic2024chiral}, as well as in $\text{SU}(N)$ Heisenberg model~\cite{hermele2009mott,chen2016synthetic,yao2021topological}. 
Therefore, our calculation and analysis capture the physics of the CSL  at intermediate coupling strength,the qualitative trend of the phase diagrams as $N$ increases
and the corresponding phase transition between IQH and CSL. 

\section{Data availability}

To facilitate the future study, the core part of the code for the parton mean-field approach and solution of the QBE equation have been uploaded for open access\cite{data}.  All data and codes are available upon request. 

\section{Acknowledgments---}XYS is grateful for related collaboration with Stefan Divic, Valentine Crepel, Tomohiro Soejima, Andrew Millis, Michael Zaletel and Ashvin Vishwanath. We appreciate discussion with Subir Sachdev, Yong Baek Kim, Ya-Hui Zhang, Michael Hermele, Gang Chen, Xuping Yao and Ying Zhong. Lu Zhang and RongNing Liu are supported by Early Career Scheme of Hong Kong Research Grant Council with grant No. 26309524 and startup fund at HKUST.

\bibliography{main}

@article{laughlin_anyon,
  title = {Superconducting Ground State of Noninteracting Particles Obeying Fractional Statistics},
  author = {Laughlin, R. B.},
  journal = {Phys. Rev. Lett.},
  volume = {60},
  issue = {25},
  pages = {2677--2680},
  numpages = {0},
  year = {1988},
  month = {Jun},
  publisher = {American Physical Society},
  doi = {10.1103/PhysRevLett.60.2677},
  url = {https://link.aps.org/doi/10.1103/PhysRevLett.60.2677}
}

@article{wenchiral,
  title = {Chiral spin states and superconductivity},
  author = {Wen, X. G. and Wilczek, Frank and Zee, A.},
  journal = {Phys. Rev. B},
  volume = {39},
  issue = {16},
  pages = {11413--11423},
  numpages = {0},
  year = {1989},
  month = {Jun},
  publisher = {American Physical Society},
  doi = {10.1103/PhysRevB.39.11413},
  url = {https://link.aps.org/doi/10.1103/PhysRevB.39.11413}
}

@article{divic2024chiral,
  title={Chiral Spin Liquid and Quantum Phase Transition in the Triangular Lattice Hofstadter-Hubbard Model},
  author={Divic, Stefan and Soejima, Tomohiro and Cr{\'e}pel, Valentin and Zaletel, Michael P and Millis, Andrew},
  journal={arXiv preprint arXiv:2406.15348},
  year={2024}
}

@online{data,
  author = {}, 
  title = {The necessary data have been uploaded to the repository},
  year = 2025,
  url = {https://github.com/WiseLu/Parton-mean-field}}

@article{zhang2021su4,
  title = {SU(4) Chiral Spin Liquid, Exciton Supersolid, and Electric Detection in Moir\'e Bilayers},
  author = {Zhang, Ya-Hui and Sheng, D. N. and Vishwanath, Ashvin},
  journal = {Phys. Rev. Lett.},
  volume = {127},
  issue = {24},
  pages = {247701},
  numpages = {6},
  year = {2021},
  month = {Dec},
  publisher = {American Physical Society},
  doi = {10.1103/PhysRevLett.127.247701},
  url = {https://link.aps.org/doi/10.1103/PhysRevLett.127.247701}
}

@article{hermele2009mott,
  title={Mott Insulators of Ultracold Fermionic Alkaline Earth Atoms: Underconstrained Magnetism and Chiral Spin Liquid},
  author={Hermele, Michael and Gurarie, Victor and Rey, Ana Maria},
  journal={Physical Review Letters},
  volume={103},
  number={13},
  pages={135301},
  year={2009},
  publisher={APS}
}

@article{yao2021topological,
  title={Topological chiral spin liquids and competing states in triangular lattice SU (N) Mott insulators},
  author={Yao, Xu-Ping and Gao, Yonghao and Chen, Gang},
  journal={Physical Review Research},
  volume={3},
  number={2},
  pages={023138},
  year={2021},
  publisher={APS}
}

@article{gorshkov2010two,
  title={Two-orbital SU (N) magnetism with ultracold alkaline-earth atoms},
  author={Gorshkov, Alexey Vyacheslavovich and Hermele, M and Gurarie, V and Xu, C and Julienne, Paul S and Ye, J and Zoller, Peter and Demler, Eugene and Lukin, Mikhail D and Rey, AM},
  journal={Nature physics},
  volume={6},
  number={4},
  pages={289--295},
  year={2010},
  publisher={Nature Publishing Group UK London}
}

@article{chester2018monopole,
  title={Monopole operators in U (1) Chern-Simons-matter theories},
  author={Chester, Shai M and Iliesiu, Luca V and Mezei, Mark and Pufu, Silviu S},
  journal={Journal of High Energy Physics},
  volume={2018},
  number={5},
  pages={1--58},
  year={2018},
  publisher={Springer}
}

@article{song_2024_phase,
  title = {Phase transitions out of quantum Hall states in moir\'e materials},
  author = {Song, Xue-Yang and Zhang, Ya-Hui and Senthil, T.},
  journal = {Phys. Rev. B},
  volume = {109},
  issue = {8},
  pages = {085143},
  numpages = {20},
  year = {2024},
  month = {Feb},
  publisher = {American Physical Society},
  doi = {10.1103/PhysRevB.109.085143},
  url = {https://link.aps.org/doi/10.1103/PhysRevB.109.085143}
}

@article{divic2024anyon,
author = {Stefan Divic  and Valentin Crepel  and Tomohiro Soejima  and Xue-Yang Song  and Andrew J. Millis  and Michael P. Zaletel  and Ashvin Vishwanath },
title = {Anyon superconductivity from topological criticality in a Hofstadterâ-Hubbard model},
journal = {Proceedings of the National Academy of Sciences},
volume = {122},
number = {33},
pages = {e2426680122},
year = {2025},
doi = {10.1073/pnas.2426680122},
URL = {https://www.pnas.org/doi/abs/10.1073/pnas.2426680122},
eprint = {https://www.pnas.org/doi/pdf/10.1073/pnas.2426680122}}

@article{ding2024particle,
  title={Particle-hole asymmetric ferromagnetism and spin textures in the triangular Hubbard-Hofstadter model},
  author={Ding, Jixun K and Yang, Luhang and Wang, Wen O and Zhu, Ziyan and Peng, Cheng and Mai, Peizhi and Huang, Edwin W and Moritz, Brian and Phillips, Philip W and Feldman, Benjamin E and others},
  journal={Physical Review X},
  volume={14},
  number={4},
  pages={041025},
  year={2024},
  publisher={APS}
}

@article{chen2016synthetic,
  title={Synthetic-gauge-field stabilization of the chiral-spin-liquid phase},
  author={Chen, Gang and Hazzard, Kaden RA and Rey, Ana Maria and Hermele, Michael},
  journal={Physical Review A},
  volume={93},
  number={6},
  pages={061601},
  year={2016},
  publisher={APS}
}

@article{anderson1973resonating,
  title={Resonating valence bonds: A new kind of insulator?},
  author={Anderson, Philip W},
  journal={Materials Research Bulletin},
  volume={8},
  number={2},
  pages={153--160},
  year={1973},
  publisher={Elsevier}
}

@article{white2007neel,
  title={Ne{\'e}l order in square and triangular lattice Heisenberg models},
  author={White, Steven R and Chernyshev, AL},
  journal={Physical review letters},
  volume={99},
  number={12},
  pages={127004},
  year={2007},
  publisher={APS}
}

@article{kuhlenkamp2024chiral,
  title={Chiral Pseudospin Liquids in Moir{\'e} Heterostructures},
  author={Kuhlenkamp, Clemens and Kadow, Wilhelm and Imamo{\u{g}}lu, Ata{\c{c}} and Knap, Michael},
  journal={Physical Review X},
  volume={14},
  number={2},
  pages={021013},
  year={2024},
  publisher={APS}
}

@article{yang2024chiral,
  title={Chiral spin liquid phase in an optical lattice at mean-field level},
  author={Yang, Jian and Liu, Xiong-Jun},
  journal={Physical Review B},
  volume={109},
  number={16},
  pages={165108},
  year={2024},
  publisher={APS}
}

@article{heinz2020state,
  title={State-dependent optical lattices for the strontium optical qubit},
  author={Heinz, Andr{\'e} and Park, Annie Jihyun and {\v{S}}anti{\'c}, Neven and Trautmann, Jan and Porsev, SG and Safronova, MS and Bloch, Immanuel and Blatt, Sebastian},
  journal={Physical review letters},
  volume={124},
  number={20},
  pages={203201},
  year={2020},
  publisher={APS}
}

@article{aidelsburger2013realization,
  title={Realization of the Hofstadter Hamiltonian with ultracold atoms in optical lattices},
  author={Aidelsburger, Monika and Atala, Marcos and Lohse, Michael and Barreiro, Julio T and Paredes, B and Bloch, Immanuel},
  journal={Physical review letters},
  volume={111},
  number={18},
  pages={185301},
  year={2013},
  publisher={APS}
}

@article{cooper2019topological,
  title={Topological bands for ultracold atoms},
  author={Cooper, NR and Dalibard, J and Spielman, IB},
  journal={Reviews of modern physics},
  volume={91},
  number={1},
  pages={015005},
  year={2019},
  publisher={APS}
}

@article{witczak2012universal,
  title={Universal transport near a quantum critical Mott transition in two dimensions},
  author={Witczak-Krempa, William and Ghaemi, Pouyan and Senthil, T and Kim, Yong Baek},
  journal={Physical Review B—Condensed Matter and Materials Physics},
  volume={86},
  number={24},
  pages={245102},
  year={2012},
  publisher={APS}
}

@article{damle1997nonzero,
  title={Nonzero-temperature transport near quantum critical points},
  author={Damle, Kedar and Sachdev, Subir},
  journal={Physical Review B},
  volume={56},
  number={14},
  pages={8714},
  year={1997},
  publisher={APS}
}

@article{sachdev1998nonzero,
  title={Nonzero-temperature transport near fractional quantum Hall critical points},
  author={Sachdev, Subir},
  journal={Physical Review B},
  volume={57},
  number={12},
  pages={7157},
  year={1998},
  publisher={APS}
}

@article{xu2022interaction,
  title={Interaction-driven metal-insulator transition with charge fractionalization},
  author={Xu, Yichen and Wu, Xiao-Chuan and Ye, Mengxing and Luo, Zhu-Xi and Jian, Chao-Ming and Xu, Cenke},
  journal={Physical Review X},
  volume={12},
  number={2},
  pages={021067},
  year={2022},
  publisher={APS}
}

@article{lee1995quantum,
  title={Quantum Boltzmann equation of composite fermions interacting with a gauge field},
  author={Lee, PA and Wen, X-G},
  journal={PHYSICAL REVIEW-SERIES B-},
  volume={52},
  pages={17--275},
  year={1995},
  publisher={AMERICAN PHYSICAL SOCIETY}
}

@book{mahan2013many,
  title={Many-particle physics},
  author={Mahan, Gerald D},
  year={2013},
  publisher={Springer Science \& Business Media}
}

@article{savary2016quantum,
  title={Quantum spin liquids: a review},
  author={Savary, Lucile and Balents, Leon},
  journal={Reports on Progress in Physics},
  volume={80},
  number={1},
  pages={016502},
  year={2016},
  publisher={IOP Publishing}
}

@article{zhou2017quantum,
  title={Quantum spin liquid states},
  author={Zhou, Yi and Kanoda, Kazushi and Ng, Tai-Kai},
  journal={Reviews of Modern Physics},
  volume={89},
  number={2},
  pages={025003},
  year={2017},
  publisher={APS}
}

@article{lee2008end,
  title={An end to the drought of quantum spin liquids},
  author={Lee, Patrick A},
  journal={Science},
  volume={321},
  number={5894},
  pages={1306--1307},
  year={2008},
  publisher={American Association for the Advancement of Science}
}

@article{balents2010spin,
  title={Spin liquids in frustrated magnets},
  author={Balents, Leon},
  journal={nature},
  volume={464},
  number={7286},
  pages={199--208},
  year={2010},
  publisher={Nature Publishing Group UK London}
}

@incollection{simon1998chern,
  title={The Chern-Simons Fermi liquid description of fractional quantum Hall states},
  author={Simon, Steven H},
  booktitle={Composite Fermions: A Unified View of the Quantum Hall Regime},
  pages={91--194},
  year={1998},
  publisher={World Scientific}
}

@phdthesis{kuhlenkamp2024aspects,
  title={Aspects and probes of strongly correlated quantum phases in two dimensions},
  author={Kuhlenkamp, Clemens},
  year={2024},
  school={Technische Universit{\"a}t M{\"u}nchen}
}

@book{auerbach2012interacting,
  title={Interacting electrons and quantum magnetism},
  author={Auerbach, Assa},
  year={2012},
  publisher={Springer Science \& Business Media}
}

@article{macdonald1988t,
  title={t U expansion for the Hubbard model},
  author={MacDonald, Allan H and Girvin, Steven M and Yoshioka, D t},
  journal={Physical Review B},
  volume={37},
  number={16},
  pages={9753},
  year={1988},
  publisher={APS}
}

@article{he2014chiral,
  title={Chiral spin liquid in a frustrated anisotropic kagome Heisenberg model},
  author={He, Yin-Chen and Sheng, DN and Chen, Yan},
  journal={Physical review letters},
  volume={112},
  number={13},
  pages={137202},
  year={2014},
  publisher={APS}
}

@article{yao2007exact,
  title={Exact chiral spin liquid with non-abelian anyons},
  author={Yao, Hong and Kivelson, Steven A},
  journal={Physical review letters},
  volume={99},
  number={24},
  pages={247203},
  year={2007},
  publisher={APS}
}

@article{schroeter2007spin,
  title={Spin Hamiltonian for which the chiral spin liquid is the exact ground state},
  author={Schroeter, Darrell F and Kapit, Eliot and Thomale, Ronny and Greiter, Martin},
  journal={Physical review letters},
  volume={99},
  number={9},
  pages={097202},
  year={2007},
  publisher={APS}
}

@article{bauer2014chiral,
  title={Chiral spin liquid and emergent anyons in a Kagome lattice Mott insulator},
  author={Bauer, Bela and Cincio, Lukasz and Keller, Brendan P and Dolfi, Michele and Vidal, Guifre and Trebst, Simon and Ludwig, Andreas WW},
  journal={Nature communications},
  volume={5},
  number={1},
  pages={5137},
  year={2014},
  publisher={Nature Publishing Group UK London}
}

@article{brinckmann2001renormalized,
  title={Renormalized mean-field theory of neutron scattering in cuprate superconductors},
  author={Brinckmann, Jan and Lee, Patrick A},
  journal={Physical Review B},
  volume={65},
  number={1},
  pages={014502},
  year={2001},
  publisher={APS}
}

@article{wen2002quantum,
  title={Quantum orders and symmetric spin liquids},
  author={Wen, Xiao-Gang},
  journal={Physical Review B},
  volume={65},
  number={16},
  pages={165113},
  year={2002},
  publisher={APS}
}

@article{wu2003exact,
  title={Exact SO (5) symmetry in the spin-3/2 fermionic system},
  author={Wu, Congjun and Hu, Jiang-ping and Zhang, Shou-cheng},
  journal={Physical Review Letters},
  volume={91},
  number={18},
  pages={186402},
  year={2003},
  publisher={APS}
}

@article{lauchli2006quadrupolar,
  title={Quadrupolar Phases of the S= 1 Bilinear-Biquadratic Heisenberg Model<? format?> on the Triangular Lattice},
  author={L{\"a}uchli, Andreas and Mila, Fr{\'e}d{\'e}ric and Penc, Karlo},
  journal={Physical review letters},
  volume={97},
  number={8},
  pages={087205},
  year={2006},
  publisher={APS}
}

@article{bauer2012three,
  title={Three-sublattice order in the SU (3) Heisenberg model on the square and triangular lattice},
  author={Bauer, Bela and Corboz, Philippe and L{\"a}uchli, Andreas M and Messio, Laura and Penc, Karlo and Troyer, Matthias and Mila, Fr{\'e}d{\'e}ric},
  journal={Physical Review B—Condensed Matter and Materials Physics},
  volume={85},
  number={12},
  pages={125116},
  year={2012},
  publisher={APS}
}

@article{hermele2011topological,
  title={Topological liquids and valence cluster states in two-dimensional SU (N) magnets},
  author={Hermele, Michael and Gurarie, Victor},
  journal={Physical Review B—Condensed Matter and Materials Physics},
  volume={84},
  number={17},
  pages={174441},
  year={2011},
  publisher={APS}
}

@article{hu2016variational,
  title={Variational Monte Carlo study of chiral spin liquid in quantum antiferromagnet on the triangular lattice},
  author={Hu, Wen-Jun and Gong, Shou-Shu and Sheng, DN},
  journal={Physical Review B},
  volume={94},
  number={7},
  pages={075131},
  year={2016},
  publisher={APS}
}

@article{wietek2017chiral,
  title={Chiral spin liquid and quantum criticality in extended S= 1 2 Heisenberg models on the triangular lattice},
  author={Wietek, Alexander and L{\"a}uchli, Andreas M},
  journal={Physical Review B},
  volume={95},
  number={3},
  pages={035141},
  year={2017},
  publisher={APS}
}

@article{honerkamp2004ultracold,
  title={Ultracold fermions and the SU (N) Hubbard model},
  author={Honerkamp, Carsten and Hofstetter, Walter},
  journal={Physical Review Letters},
  volume={92},
  number={17},
  pages={170403},
  year={2004},
  publisher={APS}
}

@article{rapp2008trionic,
  title={Trionic phase of ultracold fermions in an optical lattice: A variational study},
  author={Rapp, Akos and Hofstetter, Walter and Zar{\'a}nd, Gergely},
  journal={Physical Review B—Condensed Matter and Materials Physics},
  volume={77},
  number={14},
  pages={144520},
  year={2008},
  publisher={APS}
}

@article{chubukov1994theory,
  title={Theory of two-dimensional quantum Heisenberg antiferromagnets with a nearly critical ground state},
  author={Chubukov, Andrey V and Sachdev, Subir and Ye, Jinwu},
  journal={Physical Review B},
  volume={49},
  number={17},
  pages={11919},
  year={1994},
  publisher={APS}
}
\onecolumngrid 

\appendix

\section{$J-K$ model from the $\text{SU}(N)$ Hofstadter-Hubbard model}
\label{appendix:tJ model}

In this section we provide a pedagogical derivation of the $J-K$ model Eq\eqref{eq:JK} based on the perturbation theory~\cite{auerbach2012interacting}. Consider the Hamiltonian shown in the main text:
\begin{equation}
H = -t\sum_{\langle  i 
 j \rangle \alpha} \left( e^{-i\bar A_{ij}} c^\dagger_{i\alpha}c_{j\alpha} +\text{H.c.}\right)
 +\frac{U}{2}\sum_{i,\alpha\neq \beta}n_{i\alpha}n_{i\beta}.
 \label{eq:HubbardApp}
\end{equation}
At filling one electron each site and in the large-$U$ limit, each site is occupied by only one electron in the ground state. The double occupancy will lead to the penalty of energy $U$. Thus we treat the Hubbard term $H_{\text{int}}=\frac{U}{2} \sum_{i, \alpha \neq \beta} n_{i \alpha} n_{i \beta}$ as the unperturbed term. Then we can separate the Hilbert space into the single-occupied subspaces: 
$\mathcal S = \left\{ |n_1,n_2,\cdots\rangle\ ~| ~n_i = 1 ~~\text{for each i}  \right\}$
and multiple-occupied subspaces $\mathcal M$ with more then one electrons occupied at any site. The Hubbard term is diagonalized in this representation:
\begin{equation}
    H_{\text{int}}=\left(\begin{array}{ll}
P_{\mathcal S}H_{\text{int}}P_{\mathcal S} & ~~~~~0  \\
~~~~~0 & P_{\mathcal M}H_{\text{int}}P_{\mathcal M}  \\

\end{array}\right)
\end{equation}
where $P_{\mathcal S(\mathcal M)}$ is the projector that projects the many-body state to the subspaces $\mathcal{S}(\mathcal M)$. The energy separation between $\mathcal{S}$ and $\mathcal{M}$ is roughly to be $U$. While the kinetic terms are not diagonalized in this representation that could mix the high energy and low energy subspaces. We directly show the result of the perturbation first derived in Ref.~\cite{macdonald1988t}.
The Hamiltonian to the first order approximation is
\begin{equation}
    H_1 = P_{\mathcal S} \left[-\left(t\sum_{\langle ij\rangle} te^{-i\bar A_{ij}}c^\dagger_{i\alpha}c_{j\alpha}+\text{H.c.}\right)\right] P_{\mathcal S}.
\end{equation}
In the following derivation, we omit the notation of Hermitian conjugate H.c. for brevity but it is still counted implicitly. When the kinetic term $c^\dagger_{i\alpha}c_{j\alpha}$ acts on the single-occupied subspace $\mathcal{S}$, some sites should be vacant while the other sites are doubly occupied. Thus the kinetic term cannot survive under the projection $P_{\mathcal S}$. However, if the filling of the model is less than one electron per site, this term doesn't vanish. 
To second and third order perturbation the Hamiltonian reads
\begin{equation}
\begin{aligned}
    &~~~~~~~~~~~~~~~~H_2 =-\frac{t^2}{U} P_{\mathcal S}\left[\left(\sum_{\alpha,\langle ij\rangle}e^{-i\bar A_{ij}}c^\dagger_{i\alpha}c_{j\alpha}\right) \left(\sum_{\alpha',\langle i'j'\rangle}e^{-i\bar A_{j'i'}}c^\dagger_{j'\alpha'}c_{i'\alpha'}\right)\right]P_{\mathcal S}\\
    &H_3 =\frac{t^3}{U^2} P_{\mathcal S}\left[\left(\sum_{\alpha,\langle ij \rangle}e^{-i\bar A_{ij}}c^\dagger_{i\alpha}c_{j\alpha}\right)\left(\sum_{\alpha',\langle j'k'\rangle}e^{-i\bar A_{j'k'}}c^\dagger_{j'\alpha'}c_{k'\alpha'}\right) \left(\sum_{\alpha'' \langle k''i'' \rangle}e^{-i\bar A_{k''i''}}c^\dagger_{k''\alpha''}c_{i''\alpha''}\right)\right]P_{\mathcal S}.
    \end{aligned}
    \label{eq:H_per}
\end{equation}
The kinetic term $\sum_\alpha c_{j \alpha}^{\dagger} c_{k \alpha}$ will let one site to be doubly occupied and one site empty when acting on the state in $\mathcal{S}$. To keep the states into the subspace $\mathcal{S}$, it is required that the kinetic terms must form a loop. For $H_2$ this condition leads to $j' =j$ and $i' = i$. For $H_3$, the indices must satisfy the constraint:
\begin{equation}
k'' = k',~ j' = j,~i = i'' ~~~~\text{or}~~~~ k'' = j,~i = k',~ i'' = j' .  
\label{eq:constraint}
\end{equation}
Now the effective Hamiltonian is reduced to the form:
\begin{equation}
    H_{\text{eff}} = H_2+H_3,
\end{equation}
where $H_2, H_3$ are
\begin{equation}
\begin{aligned}
    H_2 &= -\frac{2t^2}{U}\sum_{\langle ij\rangle} P_{\mathcal S} \left[ c_{i\alpha}^\dagger c_{j \alpha}c^\dagger_{j \beta}c_{i\beta}  \right ]P_{\mathcal S}~,\\
    H_3 = \frac{6t^3}{U^2}&\sum_{\langle ijk \rangle} e^{-i\Phi_{ijk}} P_{\mathcal S} \left[c_{i\alpha}^\dagger c_{j\alpha}c_{j\beta}^\dagger c_{k\beta}c_{k\gamma}^\dagger c_{i\gamma} \right]P_{\mathcal S},\\
\end{aligned}
\label{eq:hamiltonian_effective}
\end{equation}
where the summation over $\alpha,\beta$ is implicit. The notation $\langle ij\rangle$ denotes the sites that belong to the same nearest bond and and $\langle ijk\rangle$ means that the three sites $i,j,k$ are in the same triangle. The counterclockwise or clockwise order of sites $i,j,k$ should be counted separately. For $H_2$ we have $\bar A_{ij}+\bar A_{ji} = 0$ while for $H_3$, the electron feels an A-B phase $\Phi_{\triangle} = \bar A_{12}+\bar A_{23}+\bar A_{31}$ while moving along the triangle. We identify the coupling $J = \frac{2t^2}{U}$ and $K =\frac{6t^3}{U^2}$. The factor $2$ in $H_2$ is due to the implicit hermitian conjugate of the kinetic term while the factor $6$ in $H_3$ comes from the fact that the summation satisfying the constraint \eqref{eq:constraint} in expression \eqref{eq:H_per} counts each triangle $6$ times.

The electron operator $c^\dagger$ can be expressed as $c^\dagger = f^\dagger_{i,\alpha} e^{i\theta_i}$ and $c = f_{i,\alpha} e^{-i\theta_i}$.
We define the operator
\begin{equation}
\begin{aligned}
    S_i^{\alpha \beta} &= c_{i\alpha}^\dagger c_{i \beta}\\
                       &= f_{i\alpha}^\dagger f_{i \beta}e^{i(\theta_i-\theta_i)} \\
                       &=f_{i\alpha}^\dagger f_{i \beta}
\end{aligned}
\end{equation}
The last equality is because the Hamiltonian $H_{\text{eff}}$ is defined in the subspace $\mathcal S$ and the rotor degrees of freedom $e^{i\theta_i}$ is suppressed. The Hamiltonian reads
\begin{equation}
    H_{JK} = J\sum_{\langle ij \rangle} S^{\alpha \beta}_i S^{\beta \alpha}_j + K \sum_{\langle ijk\rangle} \exp(-i\Phi_{ijk})S_i^{\alpha\beta} S_j^{\beta \gamma} S_k^{\gamma \alpha}.
    \label{eq:Heisenberg}
\end{equation}
It is exactly the Hamiltonian (\ref{eq:Heisenberg}) in the main body.

\section{The parton mean-field at strong coupling}
\label{appendix:mean-field}
In this section, we provide a brief  introduction to the parton mean-field approach and an easy way to implement calculation, which we believe is beneficial for future study. We start from the self-consistent mean-field Hamiltonian defined in the main text:
\begin{equation}
    H=-\mathcal{J} \sum_{\langle j i\rangle}\left(\chi_{j i} f_{i \alpha}^{\dagger} f_{j \alpha}+\text { H.c. }\right)+\mathcal{K} \sum_{\langle i j k\rangle}\left(e^{i \Phi_{i j k}} \chi_{i j} \chi_{j k} f_{k \alpha}^{\dagger} f_{i \alpha}+\text { H.c. }\right)-\sum_i \mu_i f_{i \alpha}^{\dagger} f_{i \alpha}.
\end{equation}
On the mean-field level the Hamiltonian is single-particle and can be diagonalized for each order parameter $\chi_{ij}$ realization.

We first choose a large super unit cell $N_x\times N_y$ and repeat it for $N_{k_x}(N_{k_y})$ times along $x(y)$ direction. The size of the super unit cell must be larger enough to capture the corresponding VBS state. For example, the size of the super unit cell must be large than $2\times 3$ to capture the VBS state when $N=3$. Then we can diagonalize the translation symmetry along $x$ and $y$ direction with quantum number $k_x$ and $k_y$ to reduce the computation resources. 
We start from random anstaz $\chi_{ij}$ and the random chemical potential $\mu_i$. Then the Hamiltonian can be easily diagonalized for each $k_x$ and $k_y$. From the wavefunction obtained, we can calculate the order parameter and the the density of spinon at each site:
\begin{equation}
\begin{aligned}
\chi_{ij} &= \sum_{\alpha}\langle f^\dagger_{i\alpha}f_{j\alpha}\rangle/N 
= \frac{1}{NN_{k_x}N_{k_y}} \sum_{\vec k,\alpha} \langle f^\dagger_{\vec k,i,\alpha}f_{\vec k,j,\alpha} \rangle e^{i\vec k \cdot \vec x_{ij}}\\
\langle n_{i} \rangle &= \sum_{\alpha}\langle f^\dagger_{i\alpha}f_{i\alpha}\rangle/N = \frac{1}{N_{k_x}N_{k_y}} \sum_{\vec k,\alpha} \langle f^\dagger_{\vec k,i,\alpha}f_{\vec k,i,\alpha} \rangle
\end{aligned}
\end{equation}
where $f_{\vec k i,\alpha}^\dagger$ is the creation operator of spinon in the $k$ space, which is the Fourier transformation of the $f_{i,\alpha}^\dagger$. Then we can tune the chemical potential such that $\mu_i \rightarrow \mu_i+\delta \mu_i$ repeatably until the density $\langle n_i \rangle$ is unit at each site. One of the crucial point is to figure out the $\delta \mu_i$ to make the density per site converge. It can be obtained by through the density-density correlation function:
\begin{equation}
    \delta \mu_{\boldsymbol{i}}=\sum_{j} C^{-1}\left(\vec x_{ij}, 0\right) (\langle n_{j}\rangle-1)
\end{equation}
where $C^{-1}(\vec x_{ij},0)$ is the inverse of zero frequency density density correlation function~\cite{hermele2011topological}. However, there are several drawbacks of this method. First, this approach is complex and not easy to implement. Second, the calculation of the correlation function cost computational resources. A more easy and straightforward way to obtain the $\delta \mu_i$ according to:
\begin{equation}
    \delta \mu_{\boldsymbol{i}} = \lambda_i (\langle n_{j}\rangle-1)
\end{equation}
where $\lambda_i$ is the parameter set arbitrarily at the beginning(It can be set unit for simplicity). Then we can update the density $\langle n_i\rangle'$ using the updated chemical potential $\mu_i+\delta \mu_i$.
If $\sum_i |\langle n_i \rangle'-1|>\sum_i |\langle n_i \rangle-1|$, which means that the density of sites doesn't converge to unit as expected, then we tune the $\lambda_i \rightarrow \alpha \lambda_i$, where $\alpha$ is set to be $0.9$. Then we repeat the procedure iteratively until the density converges. Then we move to the next calculation of the $\chi_{ij}$. The key step is to tune the chemical $\mu_i$ before the calculation of the order parameter $\mu_i$. The result converges within $300$ steps. 
We show the result of the order parameter $\chi_{ij}$ and the density $\langle n_i \rangle$ in the following.


\begin{figure*}
\centering
\includegraphics[width = 0.9\linewidth]{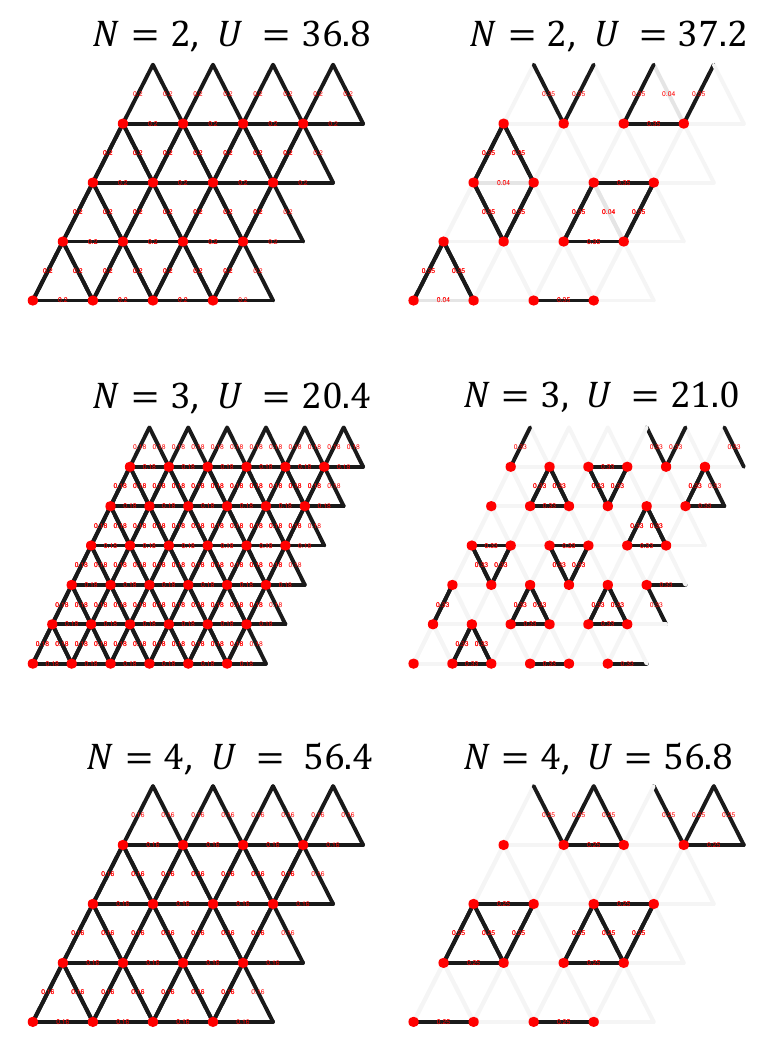}
\caption{The order parameter of CSL $\chi_{ij}$(left) and VBS(right) for different $N$ and interaction strength $U$, where we set $t = 1$. The blackness of the bond represent the absolute value of $\chi_{ij}$. The redness of the dot represent the occupation of spinon $\langle f^\dagger_{i\alpha}f_{i\alpha}\rangle/N$ at each site. Thus the occupation is $1$ at each site in all cases. For $N\geq 5$, the ground state is always chiral spin liquid.}
\label{fig:chij}
\end{figure*}

To summarize we start with a random initial ansatz, obtain the ground state at the mean-field level and calculate the occupation of spinon $\sum_\alpha\langle f_{i\alpha }^\dagger f_{i\alpha }\rangle$. Then we tune the chemical potential $\mu_i$ site by site until the number of spinon converges to $1$ at each site. We repeat the above procedure iteratively until the result converges. To make sure our result converges to the global minimum, we start from at least $50$ random initial ansatz and select the result with lowest global energy. The typical mean-field solutions of CSL and VBS  are shown in fig   \ref{fig:chij}.

\newpage

\section{Mean-field solution for the weak interaction regime}
In this section, we discuss how do we obtain the critical point between the integer quantum Hall phase and chiral spin liquid phase by solving the saddle point equation of action~\eqref{eq:partition_function}.

The $\mu_i = 0$ corresponds to $\langle L_i \rangle = 0$ at the saddle point by examining the Hamiltonian~\eqref{eq:H_slave rotor}.  With the translation symmetry on the triangular lattice, the hopping term can be diagonalized in $k$ space with eigenvalue: 
$
\xi(\vec{k})\equiv \cos{k_x}+\cos{(\frac{k_x}{2}+\frac{\sqrt{3}k_y}{2})}+\cos{(-\frac{k_x}{2}+\frac{\sqrt{3}k_y}{2})}
$, which is the single particle energy of rotor excitation.
In the CSL phase, there is no translation symmetry breaking. Thus the magnitude of the bond $\chi_{ij}\equiv t e^{-i\bar A_{ij}} \langle f_{i\alpha}^\dagger f_{j\alpha}\rangle$ should be uniform. 
Neglecting the gauge fluctuation the action of the system becomes
\begin{equation}
\begin{aligned}
     \mathcal{S} = \sum_{\vec{k}\in \text{B.Z.}}\int_\tau  \frac{1}{2U} |\partial_\tau \psi_{\vec{k}}|^2 +2\chi^f\xi(\vec{k})|\psi_{\vec{k}}|^2+  \lambda(|\psi_{\vec k}|^2-1),
\end{aligned}
\label{eq:action_rotor}
\end{equation}
where $\sum_{\vec{k}\in\text{B.Z.}}$ refers to the summation over the $\vec{k}$ in the first Brillouin zone. We assume there is no space dependence of the field $\lambda$, which is treated on a mean-field level, as the IQH and CSL phases that we focus on both preserve translation. After integrating out the rotor field $\psi$, we can obtain the saddle point equation for $\lambda$ at zero temperature~\cite{chen2016synthetic}:
\begin{equation}
    \frac{1}{N_s}\sum_{\vec{k}\in \text{B.Z.}} \frac{U}{\omega_{\vec{k}}} = 1,
    \label{eq:saddle point}
\end{equation}
where $\omega_{\vec{k}} = \sqrt{2U(\lambda-2\chi^{f}\xi(\vec{k}))}$ is the band dispersion of rotor. The vanishing energy cost of the rotor mode signals a critical instability that drives rotor condensation $\langle e^{i\theta_i} \rangle \neq 0$. This condensation restores phase coherence among the rotors, which formally identifies the  dynamics of electrons $c_{i,\alpha}$ with those of the spinons $f_{i,\alpha}$. This leads to the emergence of an IQH state as the electrons will fill Chern bands with $C=N$ which descends from the spinon Chern bands. From the Eq.~\eqref{eq:saddle point}  we can determine the critical point:
\begin{equation}\label{eq:uc1}
    U_{c_1} = 4\chi^f\left( \frac{1}{N_s}\sum_{\vec{k}\in \text{B.Z.}} \frac{1}{\sqrt{\xi(\vec{k})_{\text{max}}-\xi(\vec{k})}} \right)^{-2}.
\end{equation}
Then we make the assumption by identifying the composition of rotor and spinon hopping $\langle \psi_i^\dagger \psi_j\rangle\langle f_{i\alpha}^\dagger f_{j\alpha}\rangle$ with the electron hopping $\langle c_{i\alpha}^\dagger c_{j\alpha} \rangle$. The parameter $\chi^f$ can be estimated as $\frac{1}{zN_s}\sum_{\langle i j\rangle\alpha}te^{i\bar A_{ij}}\frac{\langle c^\dagger_{i\alpha} c_{j\alpha}\rangle}{|\langle \psi_i^\dagger \psi_j\rangle|}$ according to the definition of $\chi^f$. In the $k$ space, we have  
\begin{equation}
\chi^f \approx \frac{\bar\epsilon_k}{z|\langle \psi_i^\dagger \psi_j\rangle|}, \label{eq:J}
\end{equation}
where $|\langle \psi_i^\dagger \psi_j\rangle|$ is the expectation of hopping term of rotor, which is also translation invariant. $\bar \epsilon_k$ is the average energy of the electron, which can be calculated from the diagonalizing the single particle Hamiltonian.
By applying a Fourier transform, this spatial correlation function can be converted into the momentum-space Green function $G_\psi (\vec{k},\nu_n)$~\cite{yang2024chiral} :
\begin{equation}\label{eq:tt}
    \begin{aligned}
        |\langle \psi_i^\dagger \psi_j \rangle|  =& \frac{1}{zN_s}\sum_{\vec{k}\in \text{B.Z.}}\xi(\vec{k})\frac{1}{\beta}\sum_n G_\psi (\vec{k},\nu_n)\\
       =&\frac{1}{zN_s}\sqrt{\frac{U_{c_1}}{\chi^f}}\sum_{\vec{k}\in \text{B.Z.}}\frac{\xi(\vec{k})}{\sqrt{\xi(\vec{k})_{\text{max}}-\xi(\vec{k})}}.
    \end{aligned}
\end{equation}
Combining Eq.\ \ref{eq:uc1}, \ref{eq:J}, \ref{eq:tt} and substituting the values of $\bar\epsilon_k$, we obtain $U_{c_1}$ for different $N$.

\section{The critical transport of the rotors from quantum Boltzmann equation}
\label{appendix:qbe}
We briefly review the formalism of the quantum Boltzmann equation(QBE) describing the transport properties of the rotor for the completion of this article. The Lagrangian of the rotor coupled with the emergent gauge field near the critical point is
\begin{equation}
    \mathcal L= \frac{1}{g}|(\partial -ia)\phi|^2+i\lambda(|\phi|^2-1)+\frac{N}{4\pi}ada ,
    \label{eq:critical boson large-n}
\end{equation}
where the constraint $|\phi|^2 = 1$ is imposed by integrating out the Lagrangian multiplier $\lambda$. Usually, this cannot be exactly solved thus we resort to the RPA approximation by first solving the saddle point of $\lambda$ before considering the fluctuation around it. While, the effect of the emergent gauge field $a$ is taken account through the Chern-Simons RPA approach, present in the main text. Thus we focus on the transport of the rotor field without gauge field $a$.
The effective action of $\lambda$ can be obtained by integrating the rotor field:
\begin{equation}
    S_{\text {eff }}[\lambda]=\left[\operatorname{tr} \ln \left(-\partial^2+i \lambda\right)-\frac{i}{g} \int \lambda\right],
\end{equation}
where $\partial^2$ is the differential operator and $\text{tr}$ denotes the "trace" of the operator. However, it should be noted that the "trace" is divergent and needs to introduce the UV-cutoff $\Lambda$ to make further calculation. 
The saddle point equation of the action by variation of $\lambda$ can be obtained:

\begin{equation}
\begin{aligned}
    \frac{\delta S}{\delta \lambda} = 0& ~~~  \Rightarrow ~~~  \text{tr}\left(\frac{1}{-\partial^2+i\bar\lambda}\right) = \mathcal{A}\frac{1}{g} \\
\end{aligned}
\label{eq:lambda}
\end{equation}
where $\mathcal{A}$ is the area of the system. 
The following relations can be obtained from saddle point equation~\cite{chubukov1994theory}:
\begin{equation}
    (i\bar \lambda)^{\frac{1}{2}} = 2 T \operatorname{arcsinh}\left\{\frac{1}{2} \exp \left[-\frac{2 \pi}{T}\left(\frac{1}{g}-\frac{1}{g_c}\right)\right]\right\}.
\end{equation}
The dependence of the temperature and the deviation of the critical point is encoded in the order parameter $\bar \lambda$, which is treated as the effective mass of the rotor.

In the following, we include the fluctuation near the saddle point $\bar \lambda$.
We consider the propagator of the 
$\lambda$ field $D_\lambda$, which is obtained in the RPA approximation. In real space, the propagator of $\lambda$ field $D_\lambda(x,y)$ can be expressed as the "inverse" of current-current correlation $\langle \hat j(x) \hat j(y)\rangle$. By wick contraction, it can be obtained that
\begin{equation}
    [D_\lambda(x,y)]^{-1} =  G(x,y)G(y,x),
\end{equation}
where $G(x,y)$ is the propagator of the boson field $\phi$ and the inverse should be understood as the matrix inverse. In the operator formalism
The Green function of $\lambda$ field can be expressed as
$\hat G = \frac{1}{\partial^2 +i\bar \lambda }$.
For the convenience of the calculation we perform the Fourier transformation of the above equation. In the Fourier space we have:
 \begin{equation}
     D_\lambda^{-1}(\boldsymbol{q},i\Omega_n) = T \sum_m \int_{\boldsymbol{k}}G(\boldsymbol{k}+\boldsymbol{q},i\nu_m+i\Omega_n)G(\boldsymbol{k},i\nu_m)
 \end{equation}
The QBE can be understood in the following way: the rotor $\phi$ propagates in the space while interacting with the $\lambda$ field. The $\lambda$ field scatter with the rotor degrees of freedom.   We express the complex field $\phi$ as the composition of the creation and the annihilation operator:
\begin{equation}
    \phi(\boldsymbol{x},t) = \int_{\boldsymbol{k}} b_-(\boldsymbol{k},t)e^{i\boldsymbol{k} \cdot \boldsymbol{x}}+b_+^\dagger(\boldsymbol{k},t)e^{-i\boldsymbol{k} \cdot \boldsymbol{x}}.
\end{equation}
We define the distribution function to be
\begin{equation}
    f_{s}(t,\boldsymbol{k})=\left\langle b_{s}^{\dagger}(t, \boldsymbol{k}) b_{s}(t, \boldsymbol{k})\right\rangle,
\end{equation}
where $s$ denotes $+,-$ flavors of rotor. The quantum Boltzmann equation of the distribution function reads
\begin{equation}
\begin{aligned}
    \left(\frac{\partial}{\partial t}+s\boldsymbol{E} \frac{\partial}{\partial \boldsymbol{k}}\right)f_s(t,\boldsymbol{k}) = I_\lambda(f_{\pm}(t,\boldsymbol{k})),
\end{aligned}
\label{eq:qbe}
\end{equation}
where $I_\lambda$ are the collision terms that can be defined as
\begin{equation}
\begin{aligned}
    &I_\lambda(f_{\pm})=\int_0^\infty \frac{d\Omega}{\pi}\int\frac{d^2 \boldsymbol{q}}{(2\pi)^2}\text{Im}\left(D_{\lambda}(\boldsymbol{q},\Omega)\right) \times\\
    &\left\{
    \frac{2\pi \delta(\epsilon_{\boldsymbol{k}}-\epsilon_{\boldsymbol{k}+\boldsymbol{q}}+\Omega)}{4\epsilon_{\boldsymbol{k}}\epsilon_{\boldsymbol{k}+\boldsymbol{q}}}\left(f_s(t, \boldsymbol{k})\left[1+f_s(t, \boldsymbol{k}+\boldsymbol{q})\right] n_{\boldsymbol{q}}(\Omega)-\left[1+f_s(t, \boldsymbol{k})\right] f_s(t, \boldsymbol{k}+\boldsymbol{q})\left[1+n_{\boldsymbol{q}}(\Omega)\right]\right)\right. \\
    &\frac{2\pi \delta(\epsilon_{\boldsymbol{k}}-\epsilon_{\boldsymbol{k}+\boldsymbol{q}}-\Omega)}{4\epsilon_{\boldsymbol{k}}\epsilon_{\boldsymbol{k}+\boldsymbol{q}}}\left(f_s(t, \boldsymbol{k})\left[1+f_s(t, \boldsymbol{k}+\boldsymbol{q})\right] \left[1+n_{\boldsymbol{q}}(\Omega)\right]-\left[1+f_s(t, \boldsymbol{k})\right] f_s(t, \boldsymbol{k}+\boldsymbol{q})n_{\boldsymbol{q}}(\Omega)\right) \\
    &\left.
    \frac{2\pi \delta(\epsilon_{\boldsymbol{k}}+\epsilon_{-\boldsymbol{k}+\boldsymbol{q}}-\Omega)}{4\epsilon_{\boldsymbol{k}}\epsilon_{-\boldsymbol{k}+\boldsymbol{q}}}\left(f_s(t, \boldsymbol{k})f_{-s}(t, -\boldsymbol{k}+\boldsymbol{q})\left[1+n_{\boldsymbol{q}}(\Omega)\right] -\left[1+f_s(t, \boldsymbol{k})\right] \left[1+f_{-s}(t, -\boldsymbol{k}+\boldsymbol{q})\right]n_{\boldsymbol{q}}(\Omega)\right)
    \right\}.
\end{aligned}
\end{equation}
Here we assume that the boson of the order parameter is in the equilibrium. Thus we approximate the distribution of the boson as Bose-Einstein distribution $n_{\boldsymbol{q}}(\Omega)= 1/(e^{\beta \Omega}-1)$.  The above terms can be understood from the Fermi's-Golden rules combined with the momentum and energy conservation. 

For the convenience of the calculation we transform the Boltzmann equation into the frequency domain, with distribution function written as $f(\omega,\boldsymbol{k})$. It can be verified that the collision term vanishes if the distribution of the holon is in the equilibrium, $f(\omega,\boldsymbol{k}) = n(\omega)$. Then the Boltzmann equation can be linearized by expanding the distribution function as
\begin{equation}
    f_s(\omega,\boldsymbol{k}) = n(\omega)+s\boldsymbol{E}\cdot \boldsymbol{k} \phi(k,\omega),
\end{equation}
where $\boldsymbol{E}$ is the applied electric field and the perturbed term $\phi(k,\omega)$ only depends on the magnitude of the momentum $|\boldsymbol{k}|$.  The resulting linear differential equation can be solved numerically by finite difference method. Substituting the distribution function in Eq.~\eqref{eq:qbe} with the above leads to the linearized Boltzmann equation. This linear differential equation can be solved by finite-difference method. The detailed derivation can also be found in the Ref.~\cite{witczak2012universal}.


\end{document}